\documentclass[prl,aps,twocolumn,epsfig,superscriptaddress,showpacs]{revtex4}
\usepackage{bm}
\usepackage{amsfonts}
\usepackage[dvips]{graphicx}
\usepackage{mathrsfs}
\usepackage[intlimits]{amsmath}
\usepackage[colorlinks=false,dvipdfm]{hyperref}

\begin{document}

\author{Dong-Ling Deng}
 \affiliation{Theoretical Physics Division, Chern Institute of
Mathematics, Nankai University, Tianjin 300071, People's Republic of
China}
\author{Jing-Ling Chen}
 \email{chenjl@nankai.edu.cn}
\affiliation{Theoretical Physics Division, Chern Institute of
Mathematics, Nankai University, Tianjin 300071, People's Republic of
China}

\date{\today}

\title{Volume of Separable States for Arbitrary $N$-dimensional  System}

\begin{abstract}
In a celebrated paper ([Phys. Rev. A 58, 883 (1998)]), K.
\.{Z}yczkowski, P. Horodecki, A. Sanpera, and M. Lewenstein proved
for the first time a very interesting theorem that the volume of
separable quantum states is nonzero. Inspired by  their ideas, we
obtain a general analytical lower bound of the volume of separable
states (VOSS) for arbitrary $N$-dimensional system. Our results give
quite simple and computable sufficient conditions for separability.
Moreover, for bipartite system, an upper bound of the VOSS is also
presented.
\end{abstract}

\pacs{03.67.Mn, 03.65.Ud, 03.65.Ca} \maketitle

Quantum entanglement~\cite{A. Einstein-EPR} plays a vital role in
producing many rather nonintuitive quantum phenomena such as quantum
teleportation~\cite{C. Bennett-G.Brassard,D.Bouwmeester}, quantum
parallelism \cite{D.Deutsch}, quantum cryptographic schemes~\cite{A.
Ekert}, dense coding~\cite{C.H. Bennett-S.J.Wiesner,K. Mattle},
entanglement swapping and remote state preparation~\cite{M. A.
Nielsen}, etc.
From a formal point of view, a state is called entangled (or
inseparable) if it cannot be expressed as a mixture of product
states. Otherwise, it is separable.  Quantum entangled states can
produce many nonclassical phenomena, while separable states behave,
to some extent, more classically and cannot fulfill the task in
quantum information and computation~\cite{M.A. Nielsen-book}. One
question of great importance is then how many entangled or,
respectively, separable states there are in the set of all quantum
states. In Ref.~\cite{K.Zyczkowski-v1}, a brilliant step concerning
this problem was done for the first time. In their paper, K.
\.{Z}yczkowski \textit{et al}. proved that the set of separable
states possesses a nonzero volume under a nature measure in the
space of density matrices describing $N$-dimensional systems. This
work opened a new chapter on the study of quantum entanglement and
motivated many other works~\cite{L. Gurvits-S. J. Szarek-Guillaume}.
Nevertheless, the problem is far from been completely solved and
many questions remain open~\cite{K. Zyczkowski-1999}.


In this paper, we use a new method based on the generalized spin
matrices to estimate the volume of VOSS.
We present an analytical lower bound of the VOSS for arbitrary
$N$-dimensional multipartite system. Our results give quite simple
and computable sufficient conditions for separability. Moreover, for
bipartite system, we show a theorem that there is a neighborhood of
the maximally entangled state in which every state is entangled.
Based on this, we present a rough and approximate upper bound of the
VOSS. The results can also be generalized easily to a more general
quantum system.

To start with, we should specify some notations and definitions. Let
$H^{[N]}$ denote an $N$-dimensional Hilbert space where $
N=d_1\times d_2\times\cdots\times d_n$ and the $H_k$ ($k=1,2,\cdots,
n$) denote the $d_k$-dimensional Hilbert space of the $k$th
subsystem. A state in $H^{[N]}$ specified by a density matrix $\rho$
is said to be separable if it is a convex combination of tensor
products:
\begin{eqnarray}
\rho=\sum_{\lambda}p_{\lambda}\rho^{(1)}_{\lambda}\otimes\cdots\otimes\rho^{(n)}_{\lambda},
\end{eqnarray}
where $0\leq p_{\lambda}\leq 1$, $\sum_{\lambda}p_{\lambda}=1$ and
$\rho^{(k)}_{\lambda}$ is a density matrix on $H_k$. In Ref.\cite{A.
O. Pittenger}, A. O. Pittenger and M. H. Rubin introduced a
generalization of Pauli-spin matrices for $d$-dimensional spaces by
using of the finite Fourier transform. The generalized spin matrices
need not be hermitian but they form a basis for $d\times d$ matrices
and share many other properties with the real Pauli matrices. Here
we briefly review their methodology and main results for
completeness. Any one who want to know the details please see
Ref.\cite{A. O. Pittenger} and the reference there in.

We will begin with $d$-level systems. Let $E_{j,l}=|j\rangle\langle
l|$ denotes the computational basis of the $d$-level system and
define the adjusted basis $A=\{A_{j,l}, 0\leq j,l<d\}$ as the set of
$d\times d$ matrices defined by $A_{j,l}=E_{j,j\oplus l}$, where
$"+"$ denotes addition modulo $d$. Then the spin matrices
$S=\{S_{j,l},0\leq j,l<d\}$ are defined using the finite Fourier
transform: $(S)\equiv F(A)$. Here $F(j,l)=\texttt{Exp}(2\pi ijl/d)$.
In detail,
\begin{eqnarray}
S_{j,l}=\sum_{m=0}^{d-1}F(j,m)A_{m,l},
\end{eqnarray}
is a sum of products of scalars times matrices. Obviously, $S$ is
also a basis for the $d\times d$ matrices since $F$ is invertible.
For the $N$-dimensional Hilbert space $H^{[N]}$, the sets of
computational and adjusted bases $E^{[N]}$ for $N\times N$ matrices
are defined as:
\begin{eqnarray}
E_{\tilde{j},\tilde{l}}^{[N]}=\bigotimes_{k=1}^n
E_{j_k,l_k}^{(k)},\quad \texttt{and}\quad
A_{\tilde{j},\tilde{l}}^{[N]}=\bigotimes_{k=1}^n A_{j_k,l_k}^{(k)},
\end{eqnarray}
where $\tilde{j}$ and $\tilde{k}$ correspond to their $n$ tuples and
the superscripts in parentheses denote the  corresponding $d_k$. It
follows immediately that
$A_{\tilde{j},\tilde{l}}^{[N]}=E_{\tilde{j},\tilde{j}\oplus\tilde{l}}^{[N]}$,
where the addition of the indices is defined by:
$\tilde{j}\oplus\tilde{l}\equiv(j_1+l_1\texttt{mod} d_1, \cdots,
j_n+l_n \texttt{mod} d_n)$. Similarly, define
$F^{[N]}=F^{(1)}\otimes,\cdots,\otimes F^{(n)}$ as the usual tensor
product of the Fourier transforms $F^{(k)}$ that depend on $d_k$.
Then the corresponding set of spin matrices $S^{[N]}$ can be defined
by
$S^{[N]}_{\tilde{j},\tilde{l}}=\sum_{m=0}^{N-1}F^{[N]}(\tilde{j},\tilde{m})A^{[N]}_{\tilde{m},\tilde{l}}$,
or equivalently by
$S^{[N]}_{\tilde{j},\tilde{l}}=\bigotimes_{k=1}^n(F^{(k)}A^{(k)})_{j_k,l_k}$.
A density matrix on the $N\times N$ Hilbert space $H^{[N]}$:
$\rho^{[N]}=\sum_{\tilde{j},\tilde{l}}\rho^{[N]}_{\tilde{j},\tilde{l}}E_{\tilde{j},\tilde{l}}$
can also be expanded in adjusted bases and "spin" bases respectively
as:
\begin{eqnarray}\label{spinR}
\rho^{[N]}=\sum_{\tilde{j},\tilde{l}}a_{\tilde{j},\tilde{l}}^{[N]}A_{\tilde{j},\tilde{l}}^{[N]}
=\frac{1}{N}\sum_{\tilde{j},\tilde{l}}s_{\tilde{j},\tilde{l}}^{[N]}S_{\tilde{j},\tilde{l}}^{[N]},
\end{eqnarray}
where
$a_{\tilde{j},\tilde{l}}^{[N]}=\rho_{\tilde{j},\tilde{j}\oplus\tilde{l}}^{[N]}$
and $(s^{[N]})={F^{[N]}}^*(a^{[N]})$.

Now without proof, we  rewrite here one of the main results in
Ref.\cite{A. O. Pittenger} (Theorem $1$) as a lemma:

\textit{Lemma} $1$: If $\rho^{[N]}$ is a density matrix on
$H^{[N]}$, then $\rho^{[N]}$ is $D\equiv(d_1,\cdots,d_n)$ separable
provided
\begin{eqnarray}\label{lamma-1}
||\rho^{[N]}||_{1,D}\equiv\sum_{(\tilde{j},\tilde{l})\neq(\tilde{0},\tilde{0})}|s^{[N]}_{\tilde{j},\tilde{l}}|\leq1,
\end{eqnarray}
where $\rho^{[N]}$ has the spin representation
$\rho^{[N]}=\frac{1}{N}\sum_{\tilde{j},\tilde{l}}s_{\tilde{j},\tilde{l}}^{[N]}S_{\tilde{j},\tilde{l}}^{[N]}$
defined in term of the $D$ tensor product
$S^{[N]}_{\tilde{j},\tilde{l}}=\bigotimes_{k=1}^nS^{(k)}_{j_k,l_k}$.

Lemma $1$ provide a sufficient condition for separability of density
matrices. Our estimation of the lower bound of VOSS is based on this
condition. To process, we need another lemma:

\textit{Lemma} $2$: The matrix elements of a $N\times N$ density
matrix $\rho^{[N]}$ in the different bases satisfy the relation:
$\sum_{\tilde{j},\tilde{l}}|s^{[N]}_{\tilde{j},\tilde{l}}|^2=N\sum_{\tilde{j},\tilde{l}}|\rho^{[N]}_{\tilde{j},\tilde{l}}|^2$.

\textit{Proof}. It is very easy to prove the lemma $2$ by directly
calculation of
$\sum_{\tilde{j},\tilde{l}}|s^{[N]}_{\tilde{j},\tilde{l}}|^2$. Here
we present the key steps:
$\sum_{\tilde{j},\tilde{l}}|s^{[N]}_{\tilde{j},\tilde{l}}|^2=\texttt{Tr}[(s^{[N]})^{\dagger}(s^{[N]})]
=\texttt{Tr}[(a^{[N]})^{\dagger}({F^{[N]}}^*)^{\dagger}({F^{[N]}}^*)(a^{[N]})]=N\texttt{Tr}[(a^{[N]})^{\dagger}(a^{[N]})]=N\texttt{Tr}[{\rho^{[N]}}^2]
=N\sum_{\tilde{j},\tilde{l}}|\rho^{[N]}_{\tilde{j},\tilde{l}}|^2$.

Based on the lemma $1$ and $2$, we can obtain the lower bound of the
VOSS. Our main results are as follows:

\textit{Theorem} $1$. Let $\rho^{[N]}$ be a $N\times N$ density
matrix on $H^{[N]}$ and
$\texttt{Tr}[{\rho^{[N]}}^2]\leq\frac{N}{N^2-1}$, then $\rho^{[N]}$
is fully separable.

\textit{Proof}. On the one hand, let we calculate directly the value
of $s_{\tilde{0},\tilde{0}}^{[N]}$. Form (\ref{spinR}), it is easy
to get:
$s_{\tilde{0},\tilde{0}}^{[N]}=[{F^{[N]}}^*(a^{[N]})]_{\tilde{0},\tilde{0}}=\sum_{\tilde{l}
=\tilde{0}}^{N}{F^{[N]}_{\tilde{0},\tilde{l}}}^*a^{[N]}_{\tilde{l},\tilde{0}}=\sum_{\tilde{l}=\tilde{0}}^Na^{[N]}_{\tilde{l},\tilde{0}}
=\sum^{N}_{\tilde{l}=\tilde{0}}\rho^{[N]}_{\tilde{l},\tilde{l}}=\texttt{Tr}[\rho^{[N]}]=1$.
On the other hand, from lemma $2$, one has
$\sum_{\tilde{j},\tilde{l}}|s^{[N]}_{\tilde{j},\tilde{l}}|^2=N\sum_{\tilde{j},\tilde{l}}|\rho^{[N]}_{\tilde{j},\tilde{l}}|^2\leq\frac{N^2}{N^2-1}$.
Thus
\begin{eqnarray}
\sum_{(\tilde{j},\tilde{l})\neq(\tilde{0},\tilde{0})}|s^{[N]}_{\tilde{j},\tilde{l}}|^2\leq\frac{N^2}{N^2-1}-1=\frac{1}{N^2-1}.
\end{eqnarray}
Noting that $|s^{[N]}_{\tilde{j},\tilde{l}}|\geq0$ and using the
Lagrange Multiplier Mothods, one can easily obtain:
\begin{eqnarray}
\sum_{(\tilde{j},\tilde{l})\neq(\tilde{0},\tilde{0})}|s^{[N]}_{\tilde{j},\tilde{l}}|\leq(N^2-1)\times\frac{1}{N^2-1}=1.
\end{eqnarray}
Then from lemma $1$, we have that $\rho^{[N]}$ is fully separable.
This completes the proof of the theorem. Since
$\texttt{Tr}[{\rho^{[N]}}^2]$ is a measure of purity of state
$\rho^{[N]}$, which ranges from $\frac{1}{N}$ (for a maximally mixed
state) to $1$ (for a pure state), theorem $1$ indicate that the
purity and entanglement are closely related. All the states with
sufficiently low purity are necessarily separable.


Keeping theorem $1$ in mind and using the same method as in
Ref.\cite{K.Zyczkowski-v1} (Sec. IIIC), one can obtain the lower
bound of VOSS. However, the calculations may be tedious and the
resulting expressions very complex. To get a more distinct
expression of the lower bound, we introduce a corollary, which comes
directly from theorem $1$:

\textit{Corollary} $1$. Let $\rho'^{[N]}$ be an arbitrary $N\times
N$ density matrix on $H^{[N]}$ and
$\epsilon\leq\epsilon^*=1/\sqrt{(N^2-1)(N-1)}$ be a non-negative
real number. Then the density matrix
$\rho^{[N]}=\frac{(1-\epsilon)}{N}I+\epsilon\rho'^{[N]}$ is fully
separable on $H^{[N]}$.

\textit{Proof}. Note that $0\leq\epsilon\leq1/\sqrt{(N^2-1)(N-1)}$,
then directly calculation of $\texttt{Tr}[{\rho^{[N]}}^2]$ led to:
\begin{eqnarray}
\texttt{Tr}[{\rho^{[N]}}^2]&&=\texttt{Tr}[\frac{(1-\epsilon)^2}{N^2}I+\epsilon^2{\rho'^{[N]}}^2+2\frac{(1-\epsilon)\epsilon}{N}\rho'^{[N]}]\nonumber\\
&&\leq\frac{1+(N-1)\epsilon^2}{N}\nonumber\\
&&\leq\frac{N}{N^2-1}.
\end{eqnarray}
Then from theorem $1$, $\rho^{[N]}$ is fully separable on $H^{[N]}$
and the proof is completed. Corollary $1$ not only shows directly
that all states in the small enough neighborhood of the totally
mixed stat are separable, but also lead to the lower bound of VOSS
immediately
\begin{eqnarray}\label{loerB}
\mu(\mathcal
{S}_{sep})\geq\mu(\Delta_{\epsilon^*})=[(N^2-1)(N-1)]^{-(N-1)/2}.
\end{eqnarray}
Here, $\mu$ is the "nature measure" defined in
Ref.~\cite{K.Zyczkowski-v1}; $\mathcal {S}_{sep}$ denote the set of
separable states on $H^{[N]}$ and $\Delta_{\epsilon^*}$ is a simplex
defined as $\Delta_{\epsilon^*}=\texttt{conv}\{\mathbf{y}_i\in
\mathbb{R}^N: \mathbf{y}_i=\epsilon^*
\mathbf{x}_i+(1-\epsilon^*)\mathbf{z}_I; i=1,\cdots, N;
\mathbf{z}_I=(1/N,\cdots,1/N)\}$. Frankly speaking, the lower bound
given in (\ref{loerB}) is very rough, thus it is not better than
some previous results~\cite{L. Gurvits-S. J. Szarek-Guillaume}.
However, it successfully escapes from tedious and recondite
mathematics. What's more, it is a general analytical result and
suitable for arbitrary finite dimensional systems.

Now we have an analytical lower bound of VOSS, one may ask a natural
question: ``what is the upper bound of VOSS, or equivalently, lower
bound on the set of entangled states?" To deal with this question,
some necessary conditions for separability are needed. In
Ref.~\cite{K.Zyczkowski-v1}, K. \.{Z}yczkowski \textit{et al}.
proposed an upper bound for bipartite systems by applying the
partial transposition criterion~\cite{A. Peres}. Here, we introduce
a new method, which is based on the concurrence of multipartite
mixed states~\cite{F. Mintert,L. Aolita-2008}, to get an upper
bound. For simplicity, we only focus on two $d$-dimensional (qudit)
systems. The results can be easily generalized to a more general
system. Let $H_{AB}=H_A\otimes H_B$ denote the $d\times
d$-dimensional Hilbert space of two-qudit system, then the
concurrence for a mixed state on $H_{AB}$ is defined as the average
concurrence of the pure states of the decomposition, minimized over
all decompositions of
$\rho=\sum_{\lambda}p_{\lambda}|\psi_{\lambda}\rangle\langle\psi_{\lambda}|$:
$C(\rho)=\texttt{min}\sum_{\lambda}p_{\lambda}C(|\psi_{\lambda}\rangle)$.
Here $C(|\psi_{\lambda}\rangle)$ is the concurrence for the pure
state $|\psi_{\lambda}\rangle$ defined as: $
C(|\psi_{\lambda}\rangle)=\sqrt{2(1-\texttt{Tr}[\rho_{\lambda
A}^2])}=\sqrt{2(1-\texttt{Tr}[\rho_{\lambda B}^2])}$ with
$\rho_{\lambda
A}=\texttt{Tr}_B[|\psi_{\lambda}\rangle\langle\psi_{\lambda}|]$ be
the partial trace of $|\psi_{\lambda}\rangle\langle\psi_{\lambda}|$
over subsystem $B$ and $\rho_{\lambda B}$ a similar meaning. It is
proved in Ref.~\cite{F. Mintert} that the concurrence $C(\rho)$ has
a lower bound
\begin{eqnarray}\label{concurrence}
C^2(\rho)\geq2[\texttt{Tr}\rho^2-\texttt{Tr}\rho_A^2].
\end{eqnarray}
Inequality~(\ref{concurrence}) provides a simple sufficient
condition for entanglement. If the concurrence of a state $\rho$ is
greater than $0$, then the state is entangled. Our estimation of
lower bound on the set of inseparable states relies on the
inequality~(\ref{concurrence}).

\textit{Theorem} $2$. Let $\rho'_{AB}$ be an arbitrary density
matrix for bipartite system and $\epsilon<
\epsilon^*=\frac{d^2-\sqrt{d^4-d(d^2-1)}}{(1+d)}$ is a non-negative
number, then the density matrix
$\rho_{AB}=(1-\epsilon)|\Phi\rangle\langle\Phi|+\epsilon\rho'_{AB}$
is entangled. Here,
$|\Phi\rangle=\frac{1}{\sqrt{d}}\sum_{i=0}^{d-1}|ii\rangle$ is a
maximally entangled state on $H_{AB}$.

\textit{Proof}. We will prove this theorem by directly calculating
the lower bound of the concurrence of density matrix $\rho_{AB}$.
Form the inequality~(\ref{concurrence}), we have
\begin{eqnarray}\label{proof-Th2}
C^2&\geq&2[\texttt{Tr}\rho_{AB}^2-\texttt{Tr}\rho_A^2]\nonumber\\
&=&2\{(1-\epsilon)^2
+\texttt{Tr}[2\epsilon(1-\epsilon)\rho_{AB}'|\Psi\rangle\langle\Psi|+\epsilon^2{\rho'_{AB}}^2]\nonumber\\
&&-\texttt{Tr}[\frac{(1-\epsilon)^2}{d^2}I+\frac{2(1-\epsilon)\epsilon}{d}\rho'_A
+\epsilon^2{\rho'_A}^2]\}\nonumber\\
&\geq&2\left[(1-\epsilon)^2+\frac{\epsilon^2}{d^2}-\frac{1-\epsilon^2}{d}-\epsilon^2\right]\nonumber\\
&=&\frac{2}{d^2}[d(d-1)-2d^2\epsilon+(1+d)\epsilon^2].
\end{eqnarray}
Note that $\epsilon<\frac{d^2-\sqrt{d^4-d(d^2-1)}}{(1+d)}$, then
from inequality~(\ref{proof-Th2}), $C(\rho_{AB})>0$ is obvious.
Theorem $2$ indicate that all states in the small enough
neighborhood of the maximally entangled state $|\Psi\rangle$ are
entangled. It is interesting to note that $\epsilon^*$ is
monotonically increasing with $d$ and goes to $1/2$ as $d$ goes to
infinite. It seems like the higher the dimension, the bigger the
neighborhood of $|\Psi\rangle$ in which all states are entangled.
Theorem $2$ leads to a rough and approximate upper bound of VOSS
\begin{eqnarray}\label{upperbound}
\mu(\mathcal {S}_{sep})\leq
1-\left[\frac{d^2-\sqrt{d^4-d(d^2-1)}}{(1+d)}\right]^{N-1}.
\end{eqnarray}
Inequality~(\ref{upperbound}) only suitable for two-qudit systems,
one can generalize it to a multipartite systems using the
concurrence bound for multipartite systems in Ref.~\cite{L.
Aolita-2008}.

In summary, we have investigated the question of how many separable
or, respectively, entangled states there are in the set of all
quantum states for arbitrary $N$-dimensional systems. We present a
analytical and simple lower bound of VOSS by using the a new method
based on the separability conditions proposed in Ref.~\cite{A. O.
Pittenger}. For two-qudit system, we proved that there also exist a
neighborhood of the maximally entangled state in which all quantum
states are entangled. Based on this, we present an approximate upper
bound of the VOSS. Our results are very rough since we try to avoid
recondite mathematics and tedious calculations.

This work was supported in part by NSF of China (Grant No.
10605013), Program for New Century Excellent Talents in University,
and the Project-sponsored by SRF for ROCS, SEM.

\end{document}